\documentclass[apjl,iop,numberedappendix,twocolappendix]{emulateapj}

\usepackage{mathtools}
\usepackage{esint,graphicx,hyperref}
\hypersetup{colorlinks=false,pdfborder={0 0 0}}
\setcitestyle{notesep={; }}
\def\sgra{Sgr~A$^{\ast}$}
\def\lsim{\mathrel{\raise.3ex\hbox{$<$\kern-.75em\lower1ex\hbox{$\sim$}}}}
\def\gsim{\mathrel{\raise.3ex\hbox{$>$\kern-.75em\lower1ex\hbox{$\sim$}}}}
\defcitealias{PaperI}{Paper~I}
\defcitealias{PaperII}{Paper~II}
\defcitealias{PaperIII}{Paper~III}
\defcitealias{PaperIV}{Paper~IV}
\defcitealias{PaperV}{Paper~V}
\defcitealias{PaperVI}{Paper~VI}

\let\originalleft\left
\let\originalright\right
\renewcommand{\left}{\mathopen{}\mathclose\bgroup\originalleft}
\renewcommand{\right}{\aftergroup\egroup\originalright}
\newcommand{\ab}[1]{\left|#1\right|}
\newcommand{\br}[1]{\left[#1\right]}
\newcommand{\pa}[1]{\left(#1\right)}
\newcommand{\ed}{\mathop{}\!\mathrm{d}}

\begin{document}

\title{Universal Interferometric Signatures of a Black Hole's Photon Ring}
\shorttitle{Universal Interferometric Signatures of a Black Hole's Photon Ring}

\author{Michael~D.~Johnson\altaffilmark{1,2}, Alexandru~Lupsasca\altaffilmark{2,3,4}, Andrew~Strominger\altaffilmark{2,3}, George~N.~Wong\altaffilmark{5,6}, Shahar~Hadar\altaffilmark{2,3}, Daniel~Kapec\altaffilmark{7}, Ramesh~Narayan\altaffilmark{1,2}, Andrew~Chael\altaffilmark{1,2,8,9}, Charles~F.~Gammie\altaffilmark{5,10}, Peter~Galison\altaffilmark{2,3,11}, Daniel~C.~M.~Palumbo\altaffilmark{1}, Sheperd~S.~Doeleman\altaffilmark{1,2}, Lindy~Blackburn\altaffilmark{1,2}, Maciek~Wielgus\altaffilmark{1,2}, Dominic~W.~Pesce\altaffilmark{1,2}, Joseph~R.~Farah\altaffilmark{1,2,12}, and James~M.~Moran\altaffilmark{1}
}
\shortauthors{Michael~D.~Johnson et al.}
\altaffiltext{1}{Center for Astrophysics $|$ Harvard \& Smithsonian, 60 Garden Street, Cambridge, MA 02138, USA}
\altaffiltext{2}{Black Hole Initiative at Harvard University, 20 Garden Street, Cambridge, MA 02138, USA}
\altaffiltext{3}{Department of Physics, Harvard University, Cambridge, MA 02138, USA}
\altaffiltext{4}{Society of Fellows, Harvard University, Cambridge, MA 02138, USA}
\altaffiltext{5}{Department of Physics, University of Illinois, 1110 West Green St, Urbana, IL 61801, USA}
\altaffiltext{6}{CCS-2, Los Alamos National Laboratory, P.O. Box 1663, Los Alamos, NM 87545, USA}
\altaffiltext{7}{School of Natural Sciences, Institute for Advanced Study, Princeton, NJ 08540, USA}
\altaffiltext{8}{Princeton Center for Theoretical Science, Jadwin Hall, Princeton University, Princeton, NJ 08544, USA}
\altaffiltext{9}{NASA Hubble Fellowship Program, Einstein Fellow}
\altaffiltext{10}{Department of Astronomy, University of Illinois at Urbana-Champaign, 1002 West Green Street, Urbana, IL 61801, USA}
\altaffiltext{11}{Department of History of Science, Harvard University, Cambridge, MA 02138, USA}
\altaffiltext{12}{University of Massachusetts Boston, 100 William T.~Morrissey Blvd, Boston, MA 02125, USA}
\email{mjohnson@cfa.harvard.edu; lupsasca@fas.harvard.edu} 


\begin{abstract}
The Event Horizon Telescope image of the supermassive black hole in the galaxy M87 is dominated by a bright, unresolved ring. General relativity predicts that embedded within this image lies a thin ``photon ring," which is composed of an infinite sequence of self-similar subrings that are indexed by the number of photon orbits around the black hole. The subrings approach the edge of the black hole ``shadow," becoming exponentially narrower but weaker with increasing orbit number, with seemingly negligible contributions from high order subrings. Here, we show that these subrings produce strong and universal signatures on long interferometric baselines. These signatures offer the possibility of precise measurements of black hole mass and spin, as well as tests of general relativity, using only a sparse interferometric array.
\end{abstract}

\section{Introduction}

The EHT Collaboration has recently published images of the supermassive black hole in M87 using Very Long Baseline Interferometry (VLBI) at 1.3\,mm wavelength \citep[][hereafter Paper~I-VI]{PaperI,PaperII,PaperIII,PaperIV,PaperV,PaperVI}. These images reveal a bright ring of emission with a diameter of approximately $40\,\mu{\rm as}$. However, while the diameter of this ring is resolved by the EHT, its thickness and detailed substructure are not. In this Letter, we show that general relativity predicts an intricate substructure within this ring that presents distinctive signatures for interferometric measurements. These signatures offer a promising approach for precisely determining the masses and spins of black holes and for testing general relativity using sparse interferometers, such as an extension of the EHT to space.

\begin{figure}[t]
\centering
\includegraphics[width=.94\columnwidth]{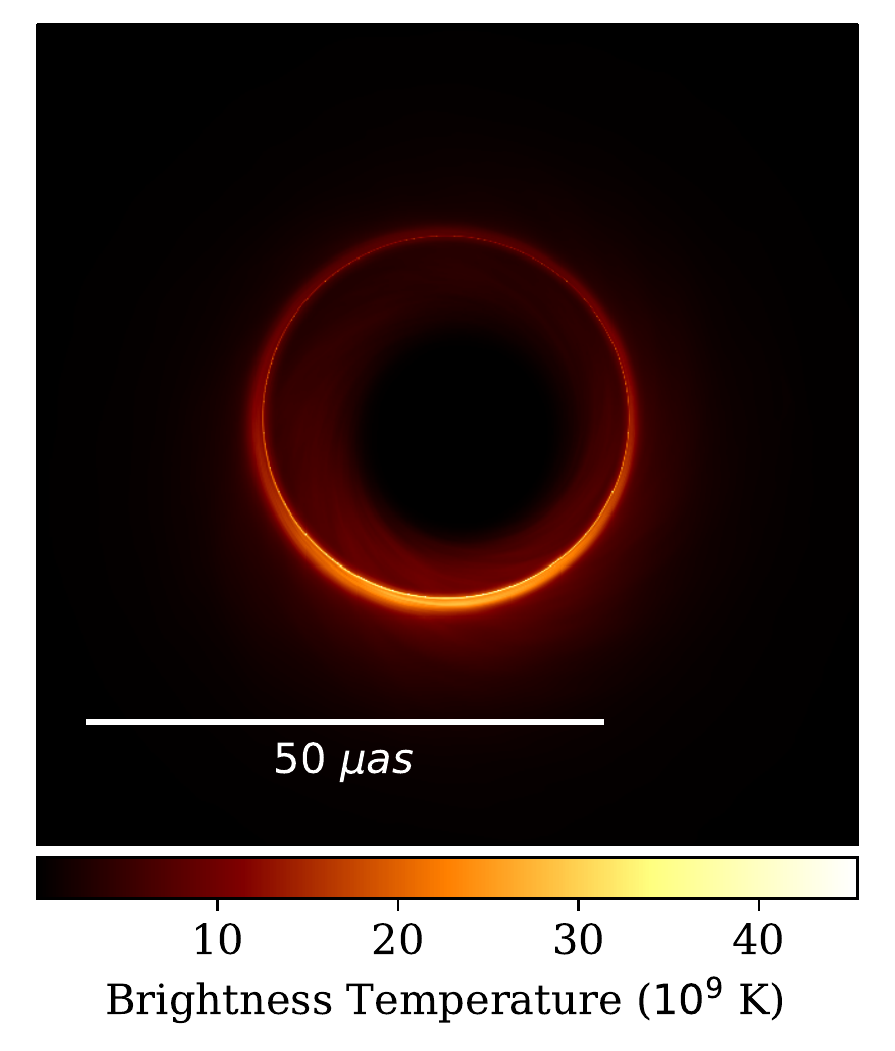}
\caption
{
Time-averaged image of a GRMHD simulation of M87 with parameters chosen to be consistent with the 2017 EHT data. This model corresponds to the high magnetic flux ``Magnetically Arrested Disk" accretion state with parameters $M=6.2\times10^9M_{\odot}$, $a/M=0.94$, $\theta_{\rm obs}=163^\circ$, $r_{\rm high}=10$, and mass accretion rate matching the 1.3\,mm flux density (see \citetalias{PaperV} for details). The spin axis points left when projected onto the image. The time average was performed over 100 snapshots produced from uniformly-spaced GRMHD fluid samples over a time range of $1000 M$ (approximately 1~year). Though visually prominent, the thin, bright ring contains only $\sim$20\% of the total image flux density.
}
\label{fig::GRMHD_Example}
\end{figure}

Neglecting opacity, a telescope with perfect resolution directed at a black hole observes an infinite number of nested images of the universe. These images arise from photons that differ by the number $n$ of half-orbits they complete around the black hole on the way from their source to the detector. Each such image is thus an increasingly delayed and demagnified snapshot of the universe as seen from the black hole. In an astrophysical setting, this self-similar sequence of relativistic images is dominated by the luminous matter surrounding the black hole and produces in its image a feature known as the  ``photon ring" of the black hole \citep{Bardeen_1973,Luminet_1979,Johannsen_Psaltis_2010,Gralla_2019}. The leading ($n=1$) subring appears as a sharp, bright feature in ray-traced images from many general-relativistic magnetohydrodynamic (GRMHD) simulations (see Figure~\ref{fig::GRMHD_Example}). Successive subrings have exponentially sharper profiles and asymptotically approach the boundary of the black hole ``shadow". For large $n$, these profiles mirror the leading subring in a manner that depends universally on the spacetime geometry, with the ratio of successive subring flux densities determined by Lyapunov exponents that characterize the instability of bound photon orbits. Hence, measuring the size, shape, and thickness of the subrings would provide new and powerful probes of a black hole spacetime.

Both GRMHD simulations and analytic estimates suggest that the photon ring should provide only $\sim$10\% of the total image flux density. This dimness may appear to preclude observations of the photon ring and its substructure, which is dimmer still. However, interferometric measurements are sensitive to more than just overall flux: they also natively filter images by their spatial wavenumbers, and therefore naturally isolate contributions from individual photon subrings. Sufficiently long baselines also resolve out diffuse flux in an image, and are thus dominated by power from the photon ring. Hence, even though sharp elements of the photon ring produce a negligible contribution to the total flux in an image, they can still provide a pronounced, dominant signal on long baselines.

In this Letter, we explore the photon ring's theoretical underpinnings and show that, surprisingly, precise measurements of the photon ring and even its subrings are feasible using interferometry. In \S\ref{sec::Theory}, we describe the shell of bound photon orbits of a Kerr black hole and its relation to the photon ring. We also present a decomposition of the photon ring into subrings indexed by half-orbit number and derive their self-similar structure, which is universally governed by Lyapunov exponents that characterize orbital instability. Next, in \S\ref{sec::Interferometric_Signatures}, we derive generically expected interferometric signatures of the photon ring. We show that its subrings produce a cascade of damped oscillations on progressively longer baselines, with the visibility of each subring conveying precise information about its diameter, width, and angular profile. Finally, in \S\ref{sec::Observational_Prospects}, we discuss observational prospects for detecting these signatures with extensions of the EHT. In particular, we highlight the possibility of detecting the leading $n=1$ subring using a station in low Earth orbit, the $n=2$ subring using a station on the Moon, and the $n=3$ subring using a station in the Sun-Earth ${\rm L}_2$ orbit.

\section{Photon Shell and Photon Ring}
\label{sec::Theory}

\begin{figure}[t]
\centering
\includegraphics[width=\columnwidth]{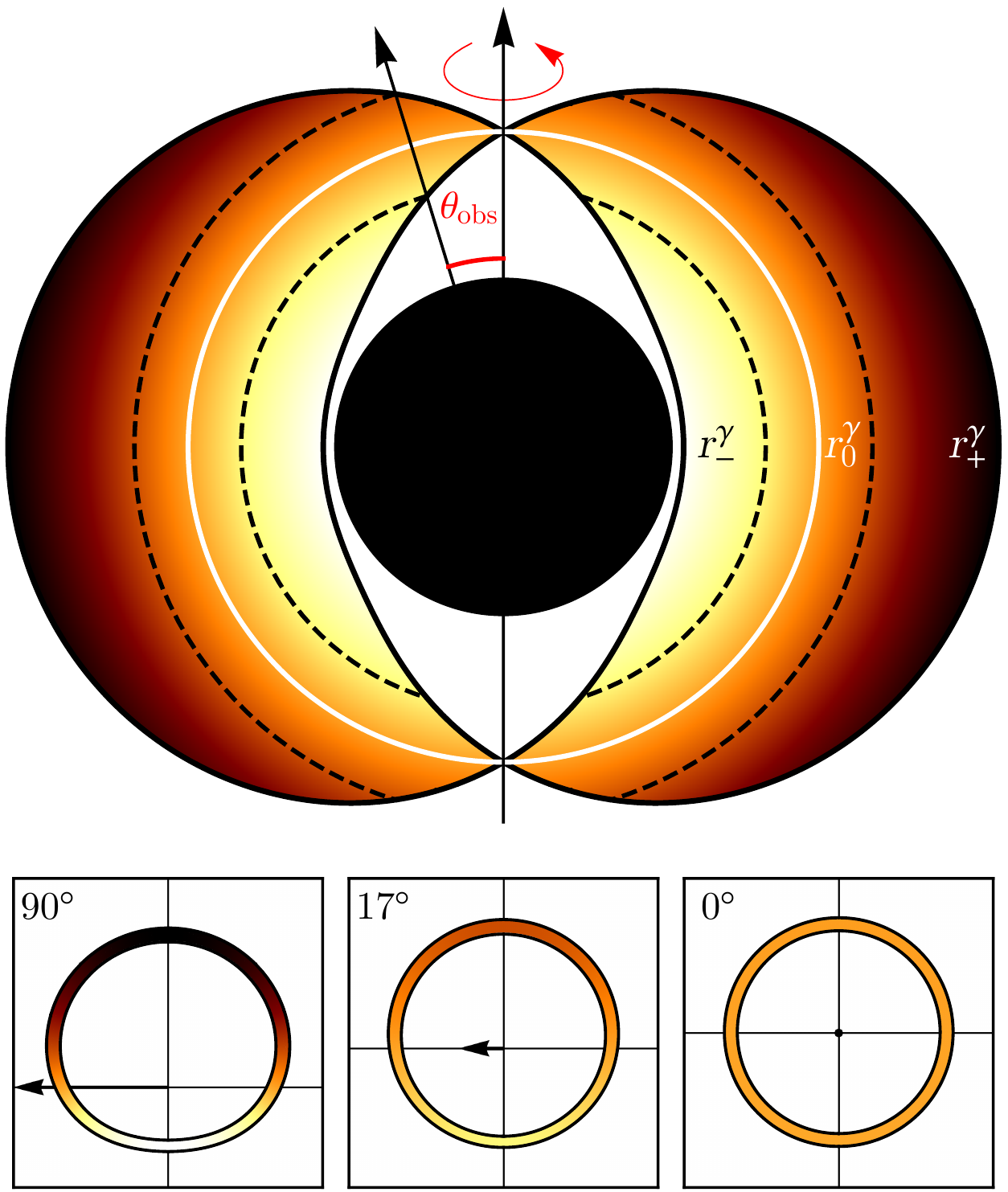}
\caption
{
Photon shell and photon ring of a Kerr black hole with spin $a/M=0.94$. Top: cross section of the photon shell in the $(r,\theta)$ plane in Boyer-Lindquist coordinates. The spin axis is vertical and the color varies with $r$. The intersection of an observer's line of sight with the photon shell boundaries at $r=r^\gamma_\pm$ determines the visible subregion of the photon shell. Bottom: photon ring on the screen of an observer at varying inclinations $\theta_{\rm obs}$ relative to the spin axis, whose projection onto the plane perpendicular to the line of sight is depicted by the (left-pointing) arrow. The center of the photon ring has a displacement from the origin that increases with spin. The color coding on the ring denotes the matching radius on the shell from which the photon emanated. The photon shell $r^\gamma_-\le r\le r^\gamma_+$ is only visible in its entirety to the edge-on ($\theta_{\rm obs}=90^\circ$) observer. The face-on ($\theta_{\rm obs}=0^\circ$) observer only receives photons from the white $r=r^\gamma_0$ orbit. The $\theta_{\rm obs}=17^\circ$ observer sees the portion of the shell delineated by the dashed lines.
}
\label{fig::Photon_Shell}
\end{figure}

This section describes the shell of unstable bound photon orbits surrounding a black hole, its lensed photon ring image, the photon subrings labeled by half-orbit number, and the angle-dependent Lyapunov exponents that govern the subring brightness ratio asymmetry. Previous treatments of these structures include \citet{Darwin_1959,Bardeen_1972,Luminet_1979,Teo_2003,Gralla_2019}. Observational aspects of these features follow in \S\ref{sec::Interferometric_Signatures} and \S\ref{sec::Observational_Prospects}.
 
\subsection{Photon Shell}

The photon shell, illustrated in Figure~\ref{fig::Photon_Shell}, is the region of a black hole spacetime containing bound null geodesics or ``bound orbits" that neither escape to infinity nor fall across the event horizon. For Schwarzschild, the photon shell is the two-dimensional sphere at $r=3M$ and any $\theta,\phi$, and $t$. For Kerr, this two-dimensional sphere fattens to a three-dimensional spherical shell. It is best described using Boyer-Lindquist coordinates, in which the metric of a Kerr black hole of mass $M$ and angular momentum $J=aM$ (with $0\le a\le M$) is
\begin{subequations}
\begin{gather}
	ds^2=-\frac{\Delta}{\Sigma}\pa{\ed t-a\sin^2{\theta}\ed\phi}^2+\frac{\Sigma}{\Delta}\ed r^2\\
	\qquad\qquad\qquad+\Sigma\ed\theta^2+\frac{\sin^2{\theta}}{\Sigma}\br{\pa{r^2+a^2}\ed\phi-a\ed t}^2,\notag\\
	\Delta=r^2-2Mr+a^2,\quad
	\Sigma=r^2+a^2\cos^2{\theta}.
\end{gather}
\end{subequations}
These coordinates have the special property that all bound orbits lie at some fixed value of $r$ in the range
\vspace{-10pt}
\begin{subequations}
\begin{align}
	r^\gamma_-&\le r\le r^\gamma_+,\\
	r^\gamma_\pm&=2M\br{1+\cos\pa{\frac{2}{3}\arccos\left(\pm\frac{a}{M}\right)}}.
\end{align}
\end{subequations}
Every point in the equatorial annulus $r^\gamma_-\le r\le r^\gamma_+$, $\theta=\pi/2$ has a unique bound orbit passing through it. On the boundaries $r=r^\gamma_\pm$, the orbits reside entirely in the equatorial plane. At generic points, on the other hand, they oscillate in the $\theta$-direction between polar angles
\begin{align}
	\label{eq::Turning_Points}
	\theta_\pm=\arccos\left(\mp\sqrt{u_+}\right),
\end{align}
where
\begin{align}
	u_\pm&=\frac{r}{a^2\pa{r-M}^2}\Big[-r^3+3M^2r-2a^2M\\
	&\qquad\pm2\sqrt{M\Delta\pa{2r^3-3Mr^2+a^2M}}\Big].\notag
\end{align}
We will refer to one such complete oscillation ($e.g.$, from $\theta_-$ back to itself) as one orbit, since the photon typically returns to a point near, but not identical to (since the azimuthal angle $\phi$ also shifts), its initial position.

To summarize, the photon shell is the spacetime region
\begin{align}
	r^\gamma_-\le r\le r^\gamma_+,\quad
	\theta_-\le\theta\le\theta_+,\quad
	0\le\phi<2\pi,
\end{align}
(depicted in Figure~\ref{fig::Photon_Shell}) for all times $-\infty\le t\le\infty$.

The bound orbit at radius $r$ has the energy-rescaled angular momentum
\begin{align}
	\label{eq::Angular_Momentum}
	\ell=\frac{M\pa{r^2-a^2}-r\Delta}{a\pa{r-M}}.
\end{align}
The inner circular equatorial orbit at $r^\gamma_-$ is prograde, while the outer one at $r^\gamma_+$ is retrograde: $\ell(r^\gamma_\mp)\gtrless0$. The overall direction of the orbits reverses at the intermediate value $r^\gamma_0$ for which $\ell$ vanishes. At that radius, $\br{\theta_-,\theta_+}$ equals $\br{0,\pi}$ and the orbits can pass over the poles.

The bound geodesics are unstable in the sense that, if perturbed slightly, they either fall into the black hole or escape to infinity where they can reach a telescope. The observed photon ring image arises from photons traveling on such ``nearly bound" geodesics. Consider two geodesics, one of which is bound, with the other initially differing only by an infinitesimal radial separation $\delta r_0$. The equation of geodesic deviation shows that, after $n$ half-orbits between $\theta_\pm$, their separation grows to
\begin{align}
	\label{eq::Geodesic_Deviation}
	\delta r_n=e^{\gamma n}\delta r_0.
\end{align} 
Here, the so-called Lyapunov exponent $\gamma$ is a function on the space of bound orbits given by\footnote{A closely related formula appears in \citet{Yang_2012}.} (see Appendix~\ref{sec::Lyapunov_Exponent})
\begin{align}
	\label{eq::Lyapunov_Exponent}
	\gamma&=\frac{4}{a}\sqrt{r^2-\frac{Mr\Delta}{\pa{r-M}^2}}\int_0^1\frac{\ed t}{\sqrt{\pa{1-t^2}\pa{u_+t^2-u_-}}}.
\end{align}
Hence, the nearly-bound geodesic will typically cross the equatorial plane a number of times of order
\begin{align}
	\label{eq::Librations}
	n\approx\frac{1}{\gamma}\ln\ab{\frac{\delta r_n}{\delta r_0}},
\end{align}
until $\delta r_n\gg\delta r_0$, when the geodesic is well-separated from the bound orbit and it shoots off to infinity (or crosses the event horizon if $\delta r_0<0$). These Lyapunov exponents are central and potentially observable quantities that characterize the geometry of the Kerr photon shell.

\subsection{Photon Ring and Subrings}
\label{sec::Photon_Ring}

The photon ring is the image on the observer screen \citep[as described by][]{Bardeen_1973} produced by photons on nearly bound geodesics. In the limit in which the photons become fully bound, it may be shown that their images approach a closed curve $C_\gamma$ given by
\begin{subequations}
\label{eq::Screen_Coordinates}
\begin{align}
	\rho&=D^{-1}\sqrt{a^2\pa{\cos^2{\theta_{\rm obs}}-u_+u_-}+\ell^2},\\
	\varphi_\rho&=\arccos\pa{-\frac{\ell}{\rho D \sin{\theta_{\rm obs}}}},
\end{align}
\end{subequations}
where $(\rho,\varphi_\rho)$ are dimensionless polar coordinates on the observer screen, while $(D,\theta_{\rm obs})$ denote the observer's distance and inclination from the Kerr spin axis. We can view $C_\gamma$ as parameterized by the shell radius $r^\gamma_-\le r\le r^\gamma_+$ from which the photon originated. For each value of $r$, \eqref{eq::Screen_Coordinates} has two solutions for $\varphi_\rho$ in the range $0\le\varphi_\rho\le 2\pi$, so each radius in the photon shell appears at two positions on $C_\gamma$. A striking consequence of \eqref{eq::Screen_Coordinates} is that for $\theta_{\rm obs}\neq0$, both $\ell$ and $\rho$, and hence $\varphi_\rho$, are functions only of $r$, $\theta_{\rm obs}$, and $D$. Hence, a measurement at a specific angle $\varphi_\rho$ along the ring probes a specific radius $r$ of the Kerr geometry and not, as might have been expected, a specific angle around the black hole!

Astrophysically observed photon intensities $I_\mathrm{ring}(\rho,\varphi_\rho)$ at the screen can be computed by backward ray-tracing. One follows the null geodesics from the observer screen back into the Kerr spacetime, integrating the Doppler-shifted strength $J$ of matter sources along the geodesic, with attenuation factors accounting for the optical depth \citep[for the images in this paper, we used {\tt ipole};][]{Moscibrodzka_2018}.\footnote{Scattering effects are negligible because the expected plasma frequency and electron gyroradius are in the MHz range, several orders of magnitude below the observing frequencies we consider.} A light ray aimed exactly at the curve $C_\gamma$ is captured by the photon shell and (unstably) orbits the black hole forever. Those aimed inside $C_\gamma$ fall into the black hole, while those aimed outside escape to infinity. Therefore, $C_\gamma$ is the edge of the black hole ``shadow".

If we shoot a light ray very near, a distance $\delta\rho$ from the shadow edge at $\rho_c$, it will circle many times through the emission region before falling into the black hole or escaping to infinity. The affine length of the ray and its number of half-orbits accordingly diverge as $\delta\rho\to0$:
\begin{align}
	\label{eq::Log_Divergence}
	n\approx-\frac{1}{\gamma}\ln\ab{\frac{\delta\rho}{\rho_c}}. 
\end{align}
This follows from \eqref{eq::Librations} together with a computed relation between $\delta\rho$ and $\delta r_0$. For optically thin matter distributions, \eqref{eq::Log_Divergence} implies a mild divergence in the observed ring intensity $I_\mathrm{ring}\sim n$ as the shadow edge is approached, since a light ray that completes $n$ half-orbits through the emission region can collect $\sim n$ times more photons along its path. The photon ring is then the bump in the photon intensity containing this logarithmic divergence at the shadow edge. Although the divergence is cut off by a finite optical depth, this striking feature remains visually prominent in many ray-traced images of GRMHD simulations, as in Figure~\ref{fig::GRMHD_Example}.

The photon ring can be subdivided into subrings arising from photons that have completed $n$ half-orbits between their source and the screen.\footnote{This definition for the photon ring agrees with that in \citet{Beckwith_2005}, but differs from the later usage in \citet{Johannsen_Psaltis_2010} and \citet{Gralla_2019} by the inclusion of the $n=1$ and 2 contributions. These low $n$ contributions fully account for the thin ring image visible in Figure~\ref{fig::GRMHD_Example}.} In order to orbit at least $n/2$ times around the black hole, the photon must be aimed within an exponentially narrowing window 
\begin{align}
	\label{eq::Window}
	\frac{\delta\rho_n}{\rho_c}\approx e^{-\gamma n}
\end{align}
around the shadow edge. Hence, the subrings occupy a sequence of exponentially nested intervals centered around $C_\gamma$. 

\begin{figure}[t]
\centering
\includegraphics[width=0.96\columnwidth]{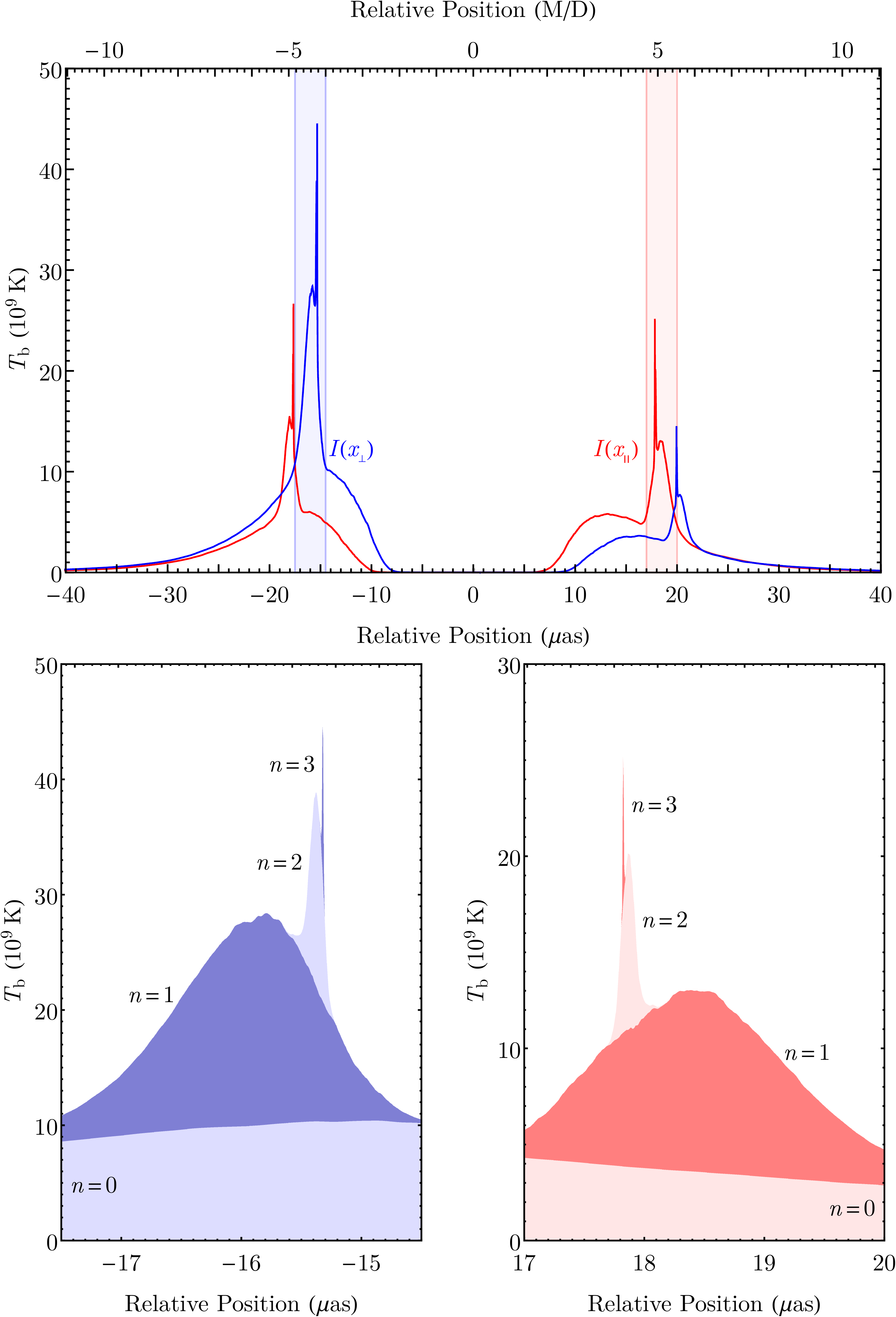}
\caption
{
Top: Brightness cross sections for the time-averaged GRMHD image shown in Figure~\ref{fig::GRMHD_Example}. The blue/red curves show cross sections perpendicular/parallel to the projected spin axis. Bottom: Decomposition of the left perpendicular peak and the right parallel peak into subrings indexed by the number $n$ of photon half-orbits executed between turning points \eqref{eq::Turning_Points} in the polar motion. Similar results are also seen in image cross sections of simple, geometrical models \citep{Gralla_2019}.
}
\label{fig::Radial_Profiles}
\end{figure}

Each subring consists of photons lensed towards the observer screen after having been collected by the photon shell from anywhere in the universe. Hence, in an idealized setting with no absorption, each subring contains a separate, exponentially demagnified image of the entire universe, with each subsequent subring capturing the visible universe at an earlier time. Together, the set of subrings are akin to the frames of a movie, capturing the history of the visible universe as seen from the black hole. In an astrophysical setting, these images are dominated by the luminous matter around the black hole. For a black hole surrounded by a uniform distribution extending over the poles, the contributions made by each subring to the total intensity profile cannot be told apart, and the individual subrings cannot be distinguished on the image. However, for a realistic disk or jet with emission peaked in a conical region, the subrings are visibly distinct: the $n^{\rm th}$ subring is approximately a smooth peak of width $e^{-\gamma n}$. Summing these smooth peaks, like layers in a tiered wedding cake (see Figure~\ref{fig::Radial_Profiles}), reproduces the leading logarithmic divergence \eqref{eq::Log_Divergence} in the intensity.

The photons comprising successive subrings for the same angle $\varphi_\rho$ traverse essentially the same orbits, and hence encounter the same matter distribution around the black hole. Apart from source variations on the timescale of an orbit, intensities of the $n^{\rm th}$ and $(n+1)^{\rm th}$ subring differ only because they correspond to windows whose widths $\delta\rho_n$ and $\delta\rho_{n+1}$ differ by a factor of $e^{-\gamma}$. Hence, for large enough $n$, the intensities are related by
\begin{align}
	\label{eq::Intensities}
	I^{n+1}_{\rm ring}(\rho_c+\delta\rho,\varphi_\rho)\approx I^n_{\rm ring}(\rho_c+e^\gamma\delta\rho,\varphi_\rho).
\end{align}
Integrating a radial slice across the subring, we therefore find the angle-dependent subring flux ratio
\begin{align}
	\label{eq::Brightness_Ratio}
	\frac{F^{n+1}_\mathrm{ring}}{F^n_\mathrm{ring}}\approx e^{-\gamma}.
\end{align}
\eqref{eq::Intensities} and \eqref{eq::Brightness_Ratio} are  matter-independent predictions for the photon ring structure that involve only general relativity. The prediction holds only for ``large enough" $n$: at small $n$, there are non-universal matter-dependent effects from photons that do not traverse exactly the same region around the black hole. Insight into when $n$ is ``large enough" might be obtained from GRMHD simulations.

Since the exponent $\gamma$ depends on $a$, $\theta_{\rm obs}$ and $\varphi_\rho$, the flux ratio asymmetry in \eqref{eq::Brightness_Ratio}  provides a new method for determination of the spin. For Schwarzschild, $\gamma=\pi$ \citep[]{Luminet_1979}, corresponding to a demagnification factor of $e^{-\pi}\approx 4\%$. For a black hole of maximal spin $a/M=1$ viewed from an inclination $\theta_{\rm obs}=17^\circ$ \citep[as estimated for M87;][]{Walker_2018}, the factor $e^{-\gamma}$ is as large as 13\% on the part of the ring where the black hole spins towards the observer. Although \eqref{eq::Window} breaks down for $n=0$, this suppression factor suggests that the leading $n=1$ subring should provide ${\sim}10\%$ of the total luminosity, in order-of-magnitude agreement with GRMHD simulations.

\section{Interferometric Signatures of a Photon Ring}
\label{sec::Interferometric_Signatures}

This section explores the response of an interferometer to the photon ring described in \S\ref{sec::Theory} through a series of increasingly refined geometrical models. We first review the interferometric signatures of an infinitesimally thin, uniform, circular ring (\S\ref{sec::Thin_Ring}). We then extend this treatment to include rings with non-uniform brightness (\S\ref{sec::Brightness_Asymmetry}), non-zero thickness (\S\ref{sec::Radial_Profiles}), and non-circular structure (\S\ref{sec::Non_Circular_Ring}). We conclude this section by discussing specific features expected for the photon ring and its subrings (\S\ref{sec::Additional_Considerations}). The prospects for experimental detection of these features are addressed in \S\ref{sec::Observational_Prospects}.

\subsection{Visibilities for a Thin, Uniform, Circular Ring}
\label{sec::Thin_Ring}

Each baseline joining two elements of an interferometer samples a complex visibility $V(\mathbf{u})$, which corresponds to a single Fourier component of the sky image $I(\mathbf{x})$ \citep{TMS}:
\begin{align}
	V(\mathbf{u})=\int I(\mathbf{x})e^{-2\pi i\mathbf{u}\cdot\mathbf{x}}\ed^2\mathbf{x}.
\end{align}
Here, $\mathbf{u}$ is the dimensionless vector baseline projected orthogonal to the line of sight and measured in units of the observation wavelength $\lambda$, while $\mathbf{x}$ is a dimensionless image coordinate measured in radians.

In terms of polar coordinates $(\rho,\varphi_\rho)$ on the observer screen \eqref{eq::Screen_Coordinates}, the image and corresponding visibility function of an infinitesimally thin, uniform, circular ring are
\vspace{-10pt}
\begin{subequations}
\begin{align}
	I(\rho,\varphi_\rho)&=\frac{1}{\pi d}\delta\pa{\rho-\frac{d}{2}},\\
	V(u,\varphi_u)&=J_0(\pi du), 
\end{align}
\end{subequations}
where $d$ is the ring diameter in radians and the image is normalized to have a total flux density of unity, $V(0)=1$. $J_m$ denotes the $m^{\rm th}$ Bessel function of the first kind, which admits the asymptotic expansion
\begin{align}
	\label{eq::Bessel_Approximation}
	J_m(\pi du)\approx\frac{1}{\pi}\sqrt{\frac{2}{du}}\cos\br{\pi\pa{du-\frac{2m+1}{4}}},
\end{align}
valid for $\pi du\gg m^2$. Hence, $V(u)$ is a weakly damped pure frequency with period $\Delta u=2/d$ inside an envelope that falls as $1/\sqrt{u}$.

\subsection{Visibilities for a Non-Uniform Ring}
\label{sec::Brightness_Asymmetry} 

The image of a thin ring with non-uniform brightness in $\varphi_\rho$ decomposes into a sum over angular Fourier modes, 
\begin{align}
    I(\rho,\varphi_\rho)=\frac{1}{\pi d}\delta\pa{\rho-\frac{d}{2}}\sum_{m=-\infty}^\infty\beta_me^{im\varphi_\rho},
\end{align}
where $\beta_{-m}=\beta_m^\ast$ since the image is real. The total image flux density is given by $\beta_0>0$. 

The corresponding visibility function is
\begin{align}
	\label{eq::Ring_Visibility}
	V(u,\varphi_u)=\sum_{m=-\infty}^\infty\beta_mJ_m(\pi du)e^{im(\varphi_u-\pi/2)}.
\end{align}
Using \eqref{eq::Bessel_Approximation}, for long baselines we may approximate
\begin{align}
	\label{eq::Asymptotic_Ring_Visibility}
	V(u,\varphi_u)&\approx\frac{\alpha_+(\varphi_u)\cos\pa{\pi du}+\alpha_-(\varphi_u)\sin\pa{\pi du}}{\sqrt{du}},\notag\\
	\alpha_\pm(\varphi_u)&\equiv\frac{1}{\pi}\sum_{m=-\infty}^\infty\beta_me^{im\br{\varphi_u+\frac{\pi}{2}\pa{m-1\pm1}}}.
\end{align}
Thus, for sufficiently long baselines, the radial visibility function of a non-uniform thin ring is determined by a single pair of weakly damped, orthogonal modes $\alpha_\pm(\varphi_u)$. Their envelope still falls as $\ab{V(u)}\sim1/\sqrt{u}$, and the modes have a common period of $\Delta u=2/d$ in complex visibilities (or $\Delta u=1/d$ in visibility amplitudes). The angular spectrum of the image $\{\beta_m\}$ is easily retrieved from the angular spectrum of the visibilities (see Appendix~\ref{sec::Angular_Spectrum}).

\begin{figure*}[t]
\centering
\includegraphics[width=0.99\textwidth]{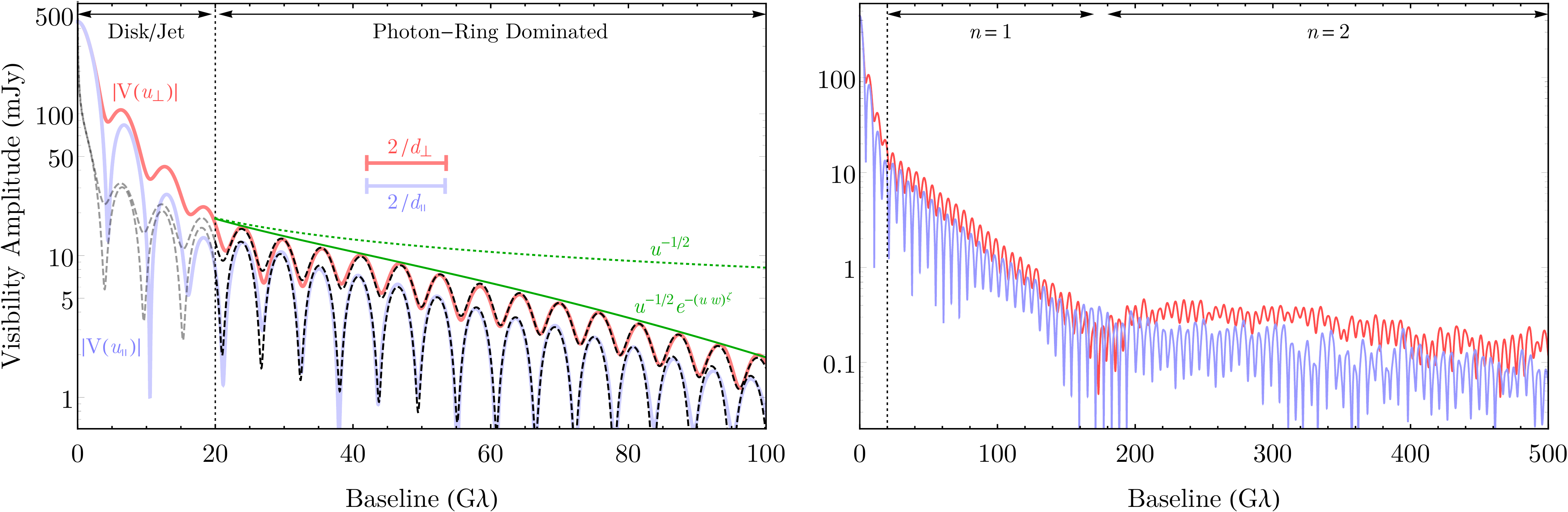}\vspace{0.2cm}
\includegraphics[width=0.99\textwidth]{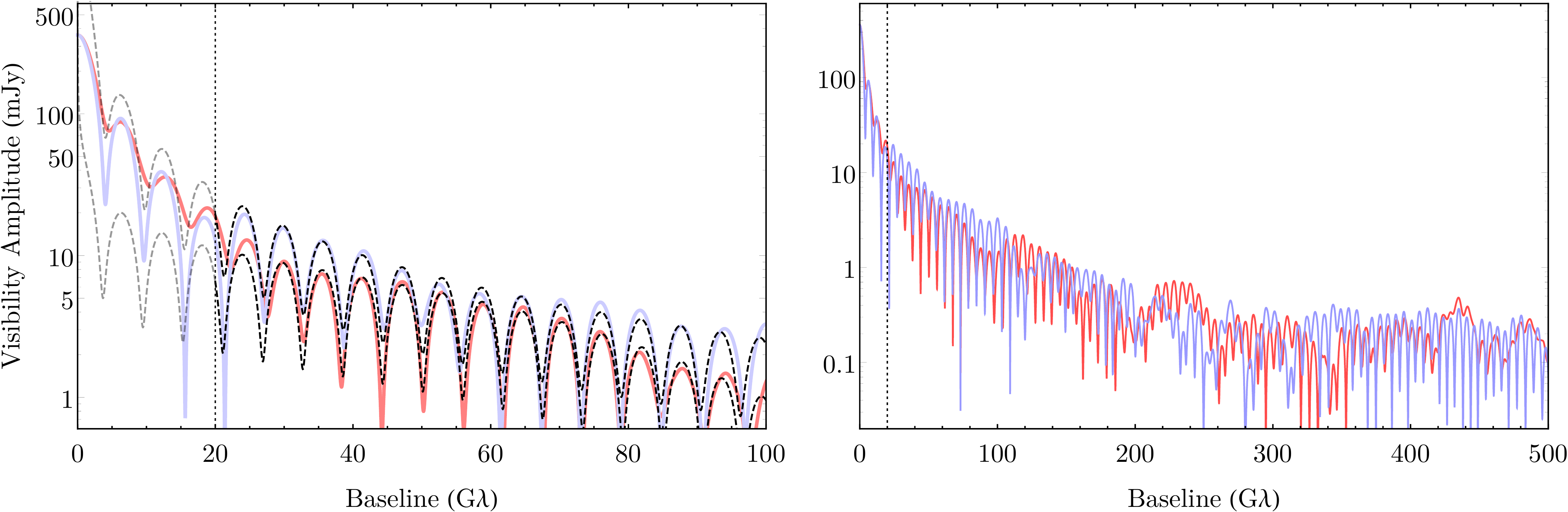}
\caption
{
Visibility amplitudes of (top) the time-averaged GRMHD simulation shown in Figure~\ref{fig::GRMHD_Example} and (bottom) a GRMHD snapshot (see Appendix~\ref{sec::GRMHD_Extra}). Amplitudes are shown for baselines perpendicular (red) and parallel (blue) to the black hole spin axis. While short baselines (left of the vertical dotted lines) display complex structure reflecting astrophysical features of the image such as emission from the disk and jet, longer baselines are dominated by the universal interferometric signatures of the photon ring. A simple model $\ab{V(u)}{=}\ab{\alpha_+\!\cos\pa{\pi du}+\alpha_-\!\sin\pa{\pi du}}\pa{du}^{-\frac{1}{2}}e^{-(wu)^\zeta}$ is overplotted (black dashed curves), with parameters determined independently along the two axes. The periodicities encode the ring diameters along each axis, and hence, $M/D$ for the black hole; their difference provides an estimate of the black hole spin and inclination. The parameters $\alpha_\pm$ carry information about the angular brightness distribution (and hence spin and inclination). The dashed green curve $u^{-1/2}$ shows the expected envelope for an infinitesimally thin ring, while the solid green curve $u^{-1/2} e^{-(wu)^\zeta}$ shows the fitted envelope that carries information about the ring thickness. On even longer baselines (right panels), the dominant visibility contributions arise from subrings with increasingly higher $n$. The universal features are more prominent in the time-averaged image, whose ring is dominated by smaller mode-numbers $m$, and which has less small-scale power outside the photon ring.
}
\label{fig::Summary_Cartoon}
\end{figure*}

\subsection{Visibilities for a Thick Ring}
\label{sec::Radial_Profiles}

Baselines of length $u\gsim1/L$ are required in order to resolve image features of size $\lsim L$. Hence, the visibility function of any ring with diameter $d$ and thickness $w\ll d$ has two asymptotic regimes:
\begin{align}
	\text{(I):}\quad\frac{1}{d}\ll u\ll\frac{1}{w},\qquad
	\text{(II):}\quad\frac{1}{d}\ll\frac{1}{w}\ll u.
\end{align}
Baselines in regime (I) resolve the diameter of the ring but not its thickness, while longer baselines in regime (II) resolve both. As such, the visibility function in regime (I) behaves like that of a thin ring (a damped periodicity with envelope $\ab{V(u)}\sim1/\sqrt{u}$), while the envelope of the visibility function in regime (II) is sensitive to the radial profile of the ring. In general, the visibility of any smooth ring decays exponentially in regime (II).\footnote{Images with discontinuous derivatives, such as a uniform disk or annulus, can have slower, power-law falloffs.}

The validity of the approximation \eqref{eq::Bessel_Approximation} in regime (I) depends on the amount of power at high values of $m$. Specifically, it requires $m_{\rm max}\lsim\sqrt{\pi d/w}$. Under this condition, $\ab{V(u)}$ has ${\sim}d/w$ periods in regime (I).

\subsection{Visibilities for a Non-Circular Ring}
\label{sec::Non_Circular_Ring}

Although the photon ring is nearly circular for all black hole spins and inclinations, the primary interferometric signatures discussed thus far do not require an image with perfectly circular structure. For instance, if an image is stretched, $I(x,y)\to I'(x,y)=I(ax,by)$, then its visibility function is correspondingly compressed, $V(u,v)\to V'(u,v)=\ab{ab}^{-1}V(u/a,v/b)$. Thus, the visibility profiles of a stretched ring share the properties and asymptotic expansions derived for a circular ring [$e.g.$, \eqref{eq::Asymptotic_Ring_Visibility}], except that the radial periodicities become a function of position angle. To leading order in the asymmetry $1-a/b$, the diameter corresponding to a damped radial periodicity in the visibility domain matches that of the stretched ring along the baseline's position angle. For the black hole in M87, the asymmetry is expected to be a few percent at most, even for a maximally rotating black hole (see Figure~\ref{fig::Image_Asymmetry}).

\subsection{Visibilities of the Photon Subrings}
\label{sec::Additional_Considerations}

As discussed in \S\ref{sec::Photon_Ring}, the photon ring decomposes into subrings labelled by the photon half-orbit number $n$. This section describes the distinctive and universal interferometric signatures of these subrings.

According to \eqref{eq::Brightness_Ratio}, the width of the radial intensity profile produced by the $n^{\rm th}$ subring is $w_n\sim w_0e^{-\gamma n}$ while the brightness remains approximately constant with $n$ (until some $n_{\rm max}$ determined by the optical depth). Each subring thus contributes a periodically modulated visibility, $V_n(u)\sim w_n/\sqrt{u}$, which falls more steeply for baselines $u>1/w_n$ (see \S\ref{sec::Radial_Profiles}). Hence, the $n^{\rm th}$ subring dominates the signal in the regime
\begin{align}
	\frac{1}{w_{n-1}}\ll u\ll\frac{1}{w_n}.
\end{align}
This implies that the totality of subring contributions has an envelope defined by this turnover behavior:
\begin{align}
	V(u)\approx\sum_{\substack{n\\w_n<1/u}}\frac{w_n}{\sqrt{u}}
	\sim\frac{1}{u^{3/2}}.
\end{align}
Together, the subrings then form a cascade of damped oscillations on progressively longer baselines, each dominated by the image of a single subring and conveying precise information about its diameter, thickness, and angular profile. Figure~\ref{fig::Summary_Cartoon} displays visibilities of the time-averaged GRMHD image in Figure~\ref{fig::GRMHD_Example}, which exhibit the expected damped periodicity, as well as clear contributions on long baselines from distinct subrings. See Figure~\ref{fig::Baseline_Lengths} for a schematic illustration of this cascade.

\section{Observational Prospects and Considerations}
\label{sec::Observational_Prospects}

\begin{figure}[t]
\centering
\includegraphics[width=0.96\columnwidth]{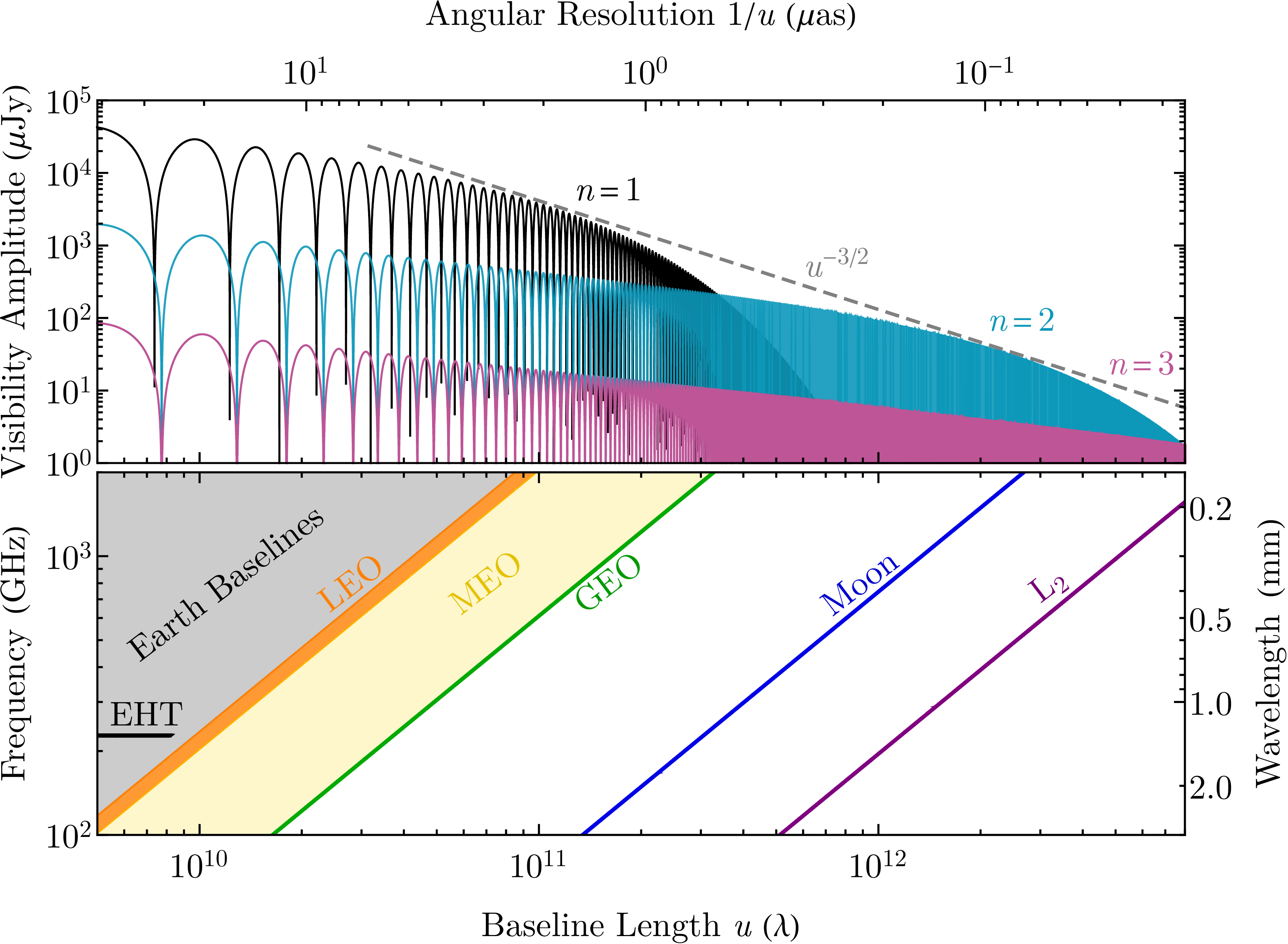}\vspace{-0.1cm}
\caption
{
Top: Schematic showing visibility amplitude as a function of baseline length for a photon ring with $d=40\,\mu{\rm as}$ and flux density comparable to M87. The black, cyan, and magenta visibilities correspond to photons with half-orbit numbers $n=1$, 2, and 3. Bottom: Frequency-dependent range of Earth baselines and representative Earth-space baselines. Earth-space baselines shown are the longest baselines for an orbiter in Low Earth Orbit (LEO), Medium Earth Orbit (MEO), geostationary orbit (GEO), on the Moon, and at the second Sun-Earth Lagrange point (${\rm L}_2$).\vspace{-0.1cm}
}
\label{fig::Baseline_Lengths}
\end{figure}

Detection of the photon ring's universal interferometric signatures requires measurements on longer baselines, with finer angular resolution than those currently available to the EHT. This extension can be achieved either by observing at higher frequencies\footnote{ALMA currently observes up to 950\,GHz (ALMA band 10) with higher frequencies (up to 1.53\,THz) possible in the future \citep[see, $e.g.$,][]{Wiedner_2007,Rigopoulou_2013}.} or by observing on longer physical baselines via space-VLBI. Figure~\ref{fig::Baseline_Lengths} shows baseline lengths for a variety of array configurations and observing frequencies.

For the EHT to observe the photon ring, it must also achieve sufficient sensitivity to detect its visibilities. For both \sgra\ and M87, the horizon-scale emission has a total flux density of $F_{\rm tot} \sim 1\,{\rm Jy}$ at $\lambda \sim 1\,{\rm mm}$ \citep[see, e.g.,][]{Bower_2015,Chael_2019}, with a fraction $f_{\rm ring} \sim 10\%$ expected to come from the photon ring (see \S\ref{sec::Theory} and \citetalias{PaperV}). The expected amplitude of the photon ring on long interferometric baselines is thus
\begin{align}
	\ab{V(\mathbf{u})}\sim30\,{\rm mJy}\pa{\frac{\ab{\mathbf{u}}}{10\,{\rm G}\lambda}}^{-3/2}&\pa{\frac{d}{40\,\mu{\rm as}}}^{-1/2}\\
	&\times\pa{\frac{f_{\rm ring}}{0.1}}\pa{\frac{F_{\rm tot}}{1\,{\rm Jy}}}.\notag
\end{align}

For comparison, a baseline from ALMA to a 4-meter orbiter with 32\,GHz of averaged bandwidth and a 10-minute coherent integration would have thermal noise of $\sigma_{\rm 950}\approx3\,{\rm mJy}$ at 950\,GHz and $\sigma_{\rm 690}\approx1.3\,{\rm mJy}$ at 690\,GHz. For baselines from ALMA to a 10-meter orbiter, such as the proposed Millimetron mission for ${\rm L}_2$ \citep{Wild_2009,Kardashev_2014}, the thermal noise would be $\sigma_{\rm 950}\approx1\,{\rm mJy}$ and $\sigma_{\rm 690}\approx0.5\,{\rm mJy}$. Another possibility would be to place a VLBI station on or orbiting the Moon, which could sample many periods of the $n=2$ regime of M87 at current EHT observing frequencies. A 10-meter dish on the Moon could achieve $\sigma\approx 0.1\,{\rm mJy}$ on baselines to ALMA with coherent integrations of 10~minutes and a bandwidth of 32\,GHz. These sensitivities could be significantly improved via simultaneous multi-frequency observations. In addition to having more sensitive receivers and longer coherence times, lower frequencies give correspondingly shorter baselines and, thus, increased interferometric power from the photon ring. Phase calibration with lower frequencies could then allow substantially longer integration times at higher frequencies. See Appendix~\ref{sec::Appendix_Observational} for additional details and discussion.

Interferometric signatures of the photon ring are most prominent when the image has little small-scale power outside the ring, and when the ring has a smooth angular profile dominated by low mode-numbers $m$. Both of these conditions are met in time-averaged images of black hole accretion flows, such as Figure~\ref{fig::GRMHD_Example}. Because visibilities of a time-averaged image are equal to time-averaged visibilities of a variable image, developing capabilities for long, coherent averaging could significantly improve the prospects for unambiguous detection and characterization of the photon ring.

In summary, precise measurements of the size, shape, thickness, and angular profile of the $n^{\rm th}$ photon subring of M87 and \sgra\ may be feasible for $n=1$ using a high-frequency ground array or low Earth orbits, for $n=2$ with a station on the Moon, and for $n=3$ with a station in ${\rm L}_2$.

\acknowledgements{We thank the National Science Foundation (AST-1440254, AST-1716536, AST-1716327, OISE-1743747, PHY-1205550) and the Gordon and Betty Moore Foundation (GBMF-5278). This work was supported in part by the Black Hole Initiative at Harvard University, which is supported by a grant from the John Templeton Foundation. 
Support for this work was provided by NASA through the NASA Hubble Fellowship grant \#HST-HF2-51431.001-A awarded by the Space Telescope Science Institute, which is operated by the Association of Universities for Research in Astronomy, Inc., for NASA, under contract NAS5-26555. 
We also thank Avery Broderick, Zachary Frankel, Jacob Goldfield, Samuel Gralla, Elizabeth Himwich, Hung-Yi Pu, and Ziri Younsi. {Author contributions:} M.D.J., A.L., A.S., S.H., R.N., A.C., P.G., and D.C.M.P.\ formulated the ideas in this paper in discussions on the theoretical aspects of the photon shell and ring. S.H., D.K., A.L., and A.S.\ developed the treatment of Lyapunov exponents that characterize the instability of bound photon orbits and determine the angle-dependent subring brightness ratios. M.D.J. conceived the idea of distinctive interferometric signatures of these subrings and derived initial expressions for them. G.N.W. adapted the {\tt ipole} software to enable high resolution images and to provide the subring decomposition from the ray tracing; he also provided the high cadence GRMHD fluid simulations.  M.D.J. and A.L. wrote the original draft; A.S., G.N.W., D.K., S.H., R.N., P.G., and D.C.M.P. provided significant text and contributions. All authors contributed to review and editing. 
}

\begin{appendix}

\section{Properties of the Photon Shell}

\begin{figure*}[t]
\centering
\includegraphics[width=\textwidth]{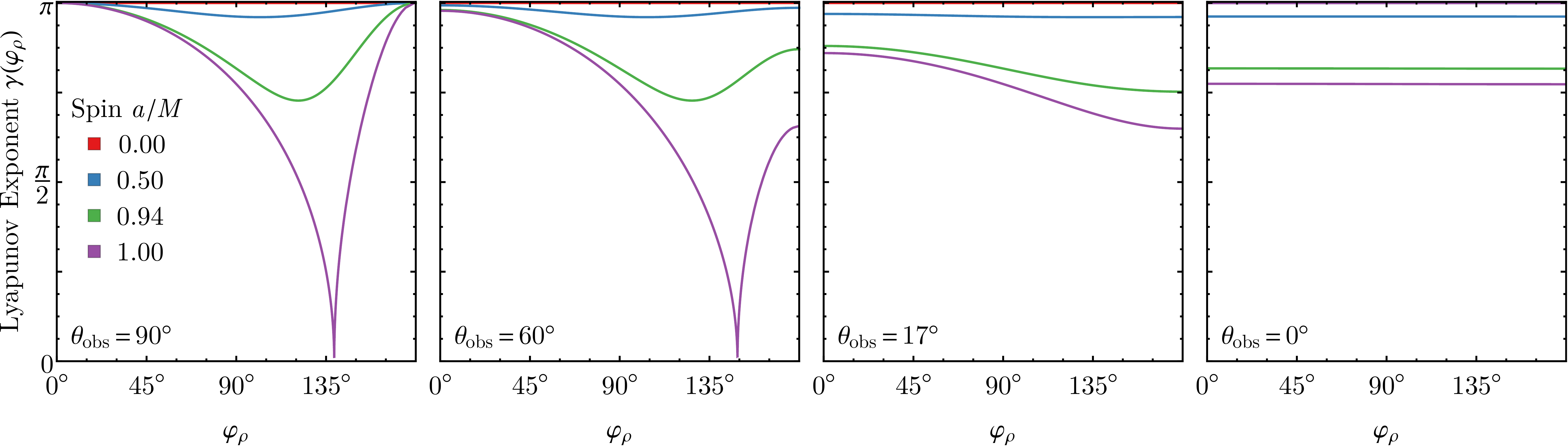}
\caption
{
Lyapunov exponent $\gamma(\varphi_\rho)$ as a function of image angle $\varphi_\rho$. The four curves show $\gamma(\varphi_\rho)$ for four spins, $a/M=0$, 0.5, 0.94, and 1, while the four panels show the results at four inclinations, $\theta_{\rm obs}=90^\circ$ (edge-on), $60^\circ$, $17^\circ$, and $0^\circ$ (face-on). Here, $\varphi_\rho$ is the angle east of north (or counterclockwise from vertical) for the lower panels in Figure~\ref{fig::Photon_Shell}; thus, the projected spin axis points toward $\varphi_\rho = 90^\circ$.
}
\label{fig::Lyapunov_Exponent}
\end{figure*}

\subsection{Geodesic Deviation From Bound Orbits}
\label{sec::Lyapunov_Exponent}

This section derives the Lyapunov exponent \eqref{eq::Lyapunov_Exponent} that governs the exponential geodesic deviation \eqref{eq::Geodesic_Deviation}, using the same conventions as \citet{Kapec_2019}. Figure~\ref{fig::Lyapunov_Exponent} shows the Lyapunov exponent $\gamma(\varphi_\rho)$ for selected values of black hole spin and inclination.

A Kerr geodesic connecting $(r_s,\theta_s)$ to $(r_o,\theta_o)$ satisfies
\begin{align}
	\fint_{r_s}^{r_o}\frac{\ed r}{\pm_r\sqrt{\mathcal{R}(r)}}=\fint_{\theta_s}^{\theta_o}\frac{\ed\theta}{\pm_\theta\sqrt{\Theta(\theta)}},
\end{align}
where $\mathcal{R}(r)$ and $\Theta(\theta)$ are potentials for the radial and polar motions, respectively, and the slash notation $\fint$ indicates that these integrals are to be evaluated along the geodesic, with turning points in each motion occurring whenever the corresponding potential vanishes.

Bound orbits occur at double roots of the radial potential where $\mathcal{R}(r)=\mathcal{R}'(r)=0$. For the bound orbit at radius $r^\gamma$, the angular integral, evaluated over one half-orbit, is the complete elliptic integral of the first kind
\begin{align}
	G_\theta(r^\gamma)\equiv\int_{\theta_-}^{\theta_+}\frac{\ed\theta}{\sqrt{\Theta(\theta)}}
	=\frac{2}{\sqrt{-u_-a^2\omega^2}}K\pa{\frac{u_+}{u_-}}.
\end{align}
Consider a nearby photon on a nearly bound geodesic, initially at a radius $r=r^\gamma+\delta r_0$ infinitesimally close to the bound orbit  ($0<\delta r_0\ll1$). After $n$ half-orbits, it advances to the larger radius $r=r^\gamma+\delta r_n$ such that
\begin{align}
	nG_\theta(r^\gamma)=\int_{r^\gamma+\delta r_0}^{r^\gamma+\delta r_n}\frac{\ed r}{\sqrt{\mathcal{R}(r)}}.
\end{align}
Since $\mathcal{R}(r^\gamma)\approx\mathcal{R}'(r^\gamma)\approx0$ for the nearly bound geodesic, the radial integral is approximately
\begin{align}
	\sqrt{\frac{2}{\mathcal{R}''(r^\gamma)}}\int_{r^\gamma+\delta r_0}^{r^\gamma+\delta r_n}\frac{\ed r}{r-r^\gamma}
	=\sqrt{\frac{2}{\mathcal{R}''(r^\gamma)}}\ln\pa{\frac{\delta r_n}{\delta r_0}}.
\end{align}
It follows that the photon's radial deviation from the bound orbit at radius $r=r^\gamma$ grows exponentially, with Lyapunov exponent
\begin{align}
	\gamma(r^\gamma)=\sqrt{\frac{\mathcal{R}''(r^\gamma)}{2}}G_\theta(r^\gamma).
\end{align}

\subsection{Edge of the Black Hole Shadow}

Inverting the cubic equation \eqref{eq::Angular_Momentum} yields the radius of the bound photon orbit with angular momentum $\ell$:
\begin{align}
	r^\gamma_\ell&=M+2M\sqrt{1-\frac{a\pa{a+\ell}}{3M^2}}\\
	&\qquad\times\cos\br{\frac{1}{3}\arccos{\frac{\pa{1-\frac{a^2}{M^2}}}{\pa{1-\frac{a\pa{a+\ell}}{3M^2}}^{3/2}}}}.\notag
\end{align}
Hence, one can also parameterize the curve $C_\gamma$ delineating the edge of the black hole shadow by $\ell$. The zero angular momentum bound orbit has radius $r^\gamma_0$.

\begin{figure}[t]
\centering
\includegraphics[width=\columnwidth]{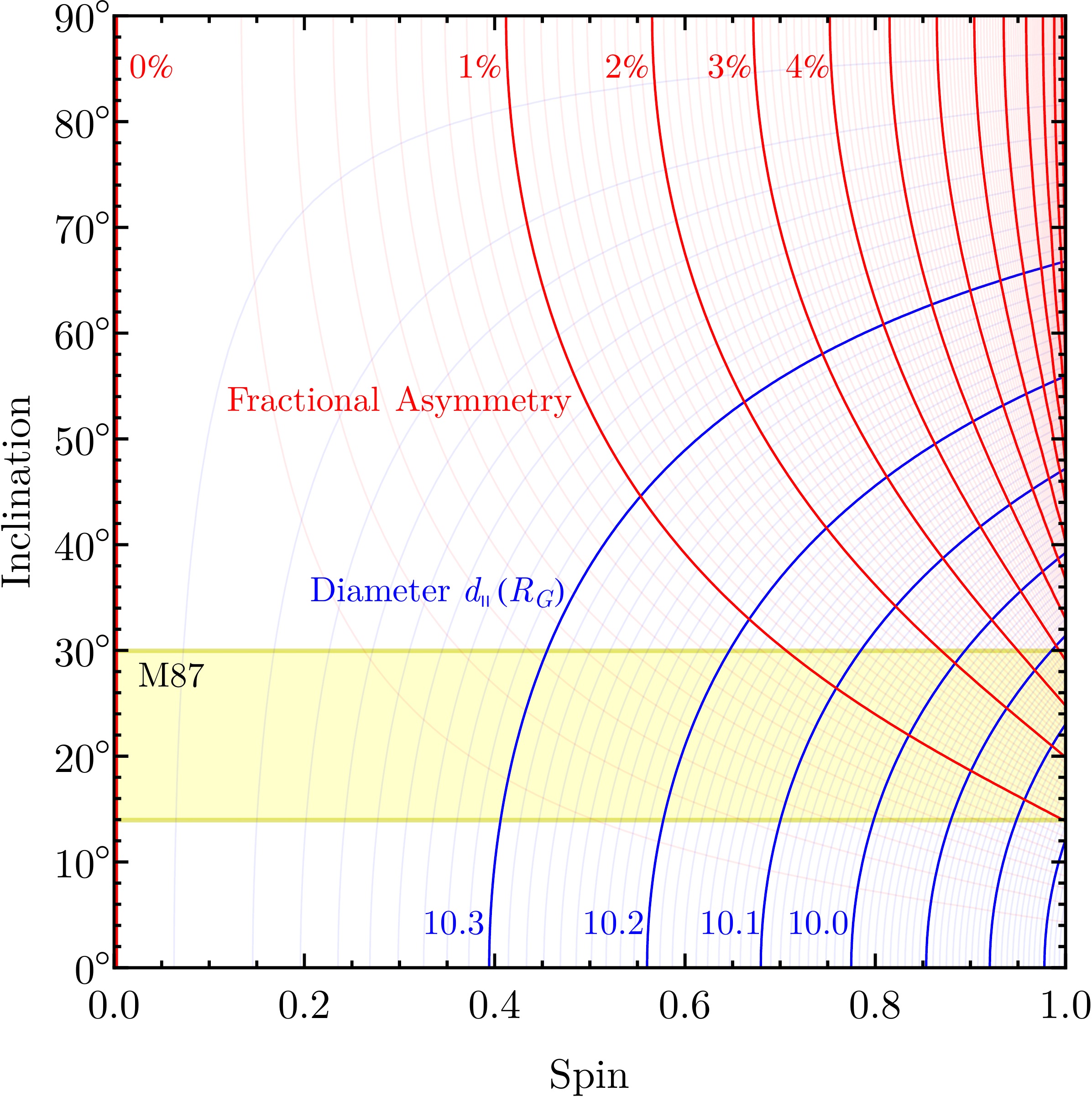}
\caption
{
Maximum black hole shadow diameter parallel to the spin axis ($d_{\parallel}$) and fractional asymmetry ($A\equiv1-d_\perp/d_\parallel$) as a function of black hole spin and inclination \citep[see also][]{Chan_2013}. Plotted contours for each are equally spaced. The shaded region shows the plausible range of inclination for M87 assuming that the black hole is aligned with the its jet \citep[see, $e.g.$,][]{Mertens_2016}. The GRMHD simulations used in this paper have $d_\parallel\approx9.837M/D$ and $d_\perp\approx9.737M/D$, giving $A=1.02\%$.
}
\label{fig::Image_Asymmetry}
\end{figure}

At large $n$, the photon subrings rapidly converge to the curve $C_\gamma$ delineating the edge of the black hole shadow. Figure~\ref{fig::Image_Asymmetry} shows the shadow size and asymmetry as a function of black hole spin and inclination. To define these quantities, we use the maximum diameters of the shadow parallel and perpendicular to the black hole spin axis, $d_{\parallel}$ and $d_{\perp}$. Measurements of non-zero asymmetry provide a lower bound on inclination and spin. When paired with complementary measurements of black hole mass, the combination of diameter and asymmetry uniquely determines the spin and inclination. More generally, measuring the full shape of the shadow uniquely determines the mass, spin, and inclination, apart from the degenerate case for face-on viewing (which gives a circular shadow), even without complementary measurements of mass.

\subsection{Subring Flux Ratio}

This section provides a more detailed derivation of \eqref{eq::Intensities}, from which \eqref{eq::Brightness_Ratio} follows, for the case of images appearing outside the black hole shadow.

Assume a geometrically and optically thin, axisymmetric, equatorial emission region around the black hole, and consider two photons with a radial turning point along their trajectory: photon (a), which impinges upon the observer screen at $(\rho_c+\delta\rho,\varphi_\rho)$ and underwent $n+1$ half-orbits after emission at radius $r^\gamma+\delta r^{(a)}_{\rm e}$, and photon (b), which arrives at $(\rho_c+e^\gamma\delta\rho,\varphi_\rho)$ after $n$ half-orbits following emission at radius $r^\gamma+\delta r^{(b)}_{\mathrm{e}}$. The radial separation $\delta r_{\rm min}$ of the turning point from the critical radius $r^\gamma$ scales like $\delta r_{\rm min}^2\sim\delta\rho$ \citep{Gralla_2019}. Therefore, $\delta r_{\rm min}^{(a)}\approx e^{-\gamma/2}\delta r_{\rm min}^{(b)}$. Photon (a) undergoes $(n+1)/2$ half orbits between $\delta r_{\rm min}^{(a)}$ and $\delta r_\mathrm{e}^{(a)}$, while photon (b) undergoes $n/2$ half orbits between $\delta r_{\rm min}^{(b)}$ and $\delta r_\mathrm{e}^{(b)}$. By \eqref{eq::Geodesic_Deviation}, it then follows that $\delta r_\mathrm{e}^{(a)}\approx\delta r_\mathrm{e}^{(b)}$. Since by assumption, the localized emissivity depends only on $r$, this shows that photons (a) and (b) arrive at the screen with approximately the same intensity, which implies \eqref{eq::Intensities}. Note that this argument relies on \eqref{eq::Geodesic_Deviation}, which holds only as long as the emission radii $\delta r_\mathrm{e}^{(a/b)}$ are close enough to $r^\gamma$.

\section{Interferometric Signatures of Rings}

This section extends the discussion of interferometric signatures of photon rings from \S\ref{sec::Interferometric_Signatures}.

\subsection{Image Centroid Displacement}

The analysis in \S\ref{sec::Interferometric_Signatures} focused on rings centered at the origin. Even neglecting measurement errors, a non-zero black hole spin causes the shadow centroid to be displaced from the origin (see Figure~\ref{fig::Photon_Shell}). However, a centroid displacement has a simple visibility signature: visibilities for an image displaced by $\Delta\mathbf{x}$ are equal to those of the original image multiplied by a linear phase slope, $e^{-2\pi i\Delta\mathbf{x}\cdot\mathbf{u}}$. This term does not modify visibility amplitudes and could be removed for experiments designed with sensitivity to absolute interferometric phase.

\begin{figure*}[t]
\centering
\hspace{0.5cm}\includegraphics[width=.96\textwidth]{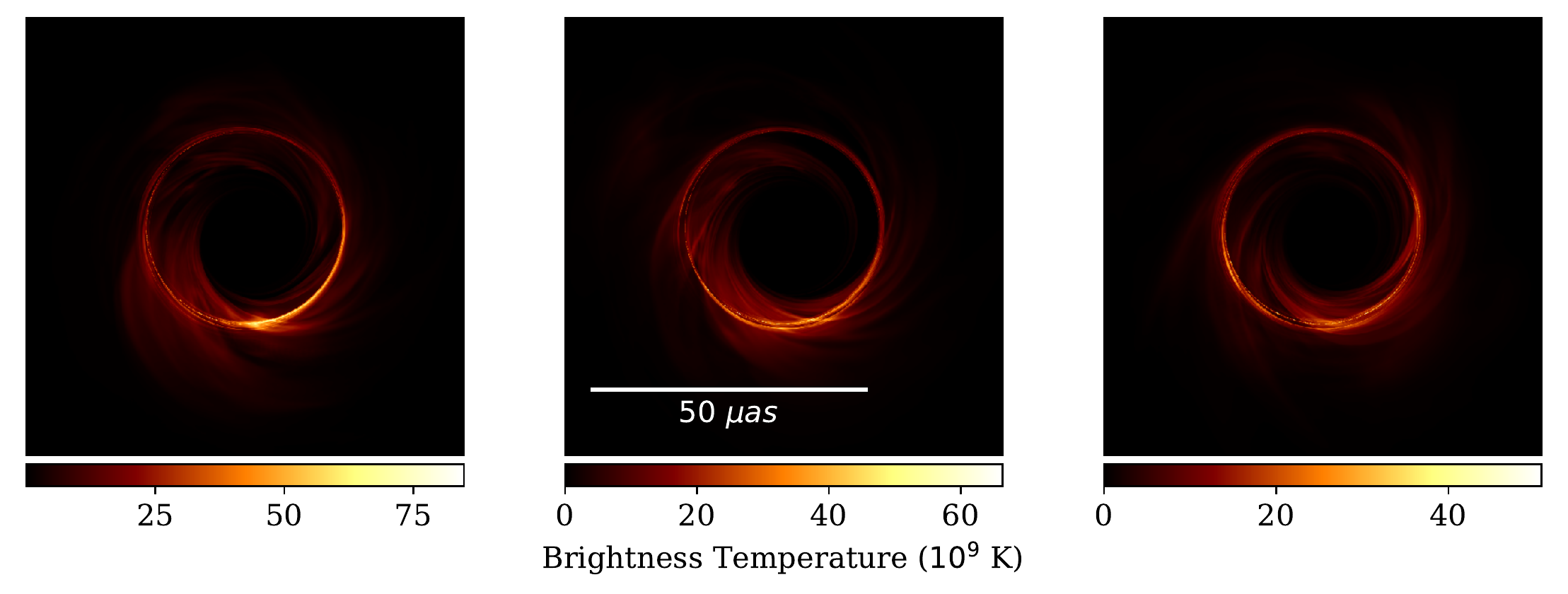}
\includegraphics[width=\textwidth]{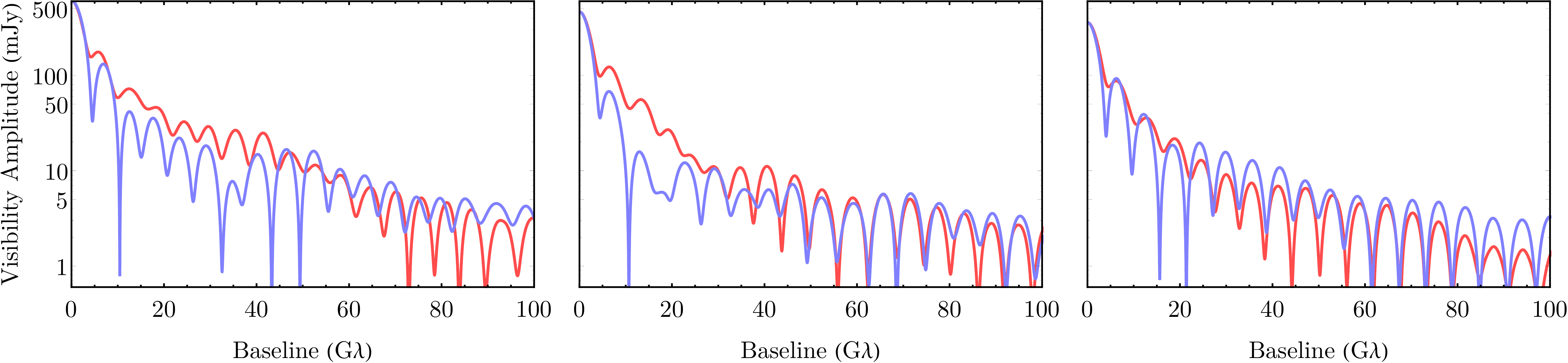}
\caption
{
Three instantaneous snapshots for the GRMHD simulation shown in Figure~\ref{fig::GRMHD_Example} (top) and their corresponding visibility amplitudes (bottom). Interferometric signatures for the snapshots are qualitatively similar to those for the time-averaged image; all show a damped radial periodicity. However, the snapshot images have more complex azimuthal structure in the photon ring as well as more small-scale structure outside the photon ring, with both effects contributing to the more complex interferometric signature on long baselines (see also Figure~\ref{fig::Summary_Cartoon}). Consequently, interferometric studies of the photon ring with a sparse interferometer will be more difficult with short observations than with long observations (${\gg}\,M$) that can be coherently combined.
}
\label{fig::Snapshots}
\end{figure*}

\subsection{Estimating the Angular Spectrum of a Ring}
\label{sec::Angular_Spectrum}

The inverse of \eqref{eq::Ring_Visibility} is\footnote{This relationship is not restricted to a ring and trivially generalizes to any image that is separable, $I(\rho,\varphi_\rho)=I_\rho(\rho)I_\varphi(\varphi_\rho)$.}
\begin{align}
	\int_0^{2\pi}V(u,\theta_u)e^{-im\theta_u}\ed\theta_u=2\pi(-i)^m\beta_mJ_m(\pi du).
\end{align}
Hence, as asserted in \S\ref{sec::Brightness_Asymmetry}, the angular image spectrum $\{\beta_m\}$ of a ring can be determined from its angular visibility spectrum. On long baselines, the asymptotic approximation for the Bessel function gives
\begin{align}
	\beta_m\approx\left\langle\sqrt{\frac{du}{2}}i^m\int_0^{2\pi}V(u,\varphi_u)e^{-im\varphi_u}\ed\varphi_u\right\rangle,
\end{align}
where $\langle\cdots\rangle$ replaces the oscillating radial component by its envelope, $\cos\br{\pi\pa{du-\frac{2m+1}{4}}}\rightarrow1$. 

\subsection{Effects from Transient Compact Structures}
\label{sec::GRMHD_Extra}

As noted in the text, interferometric signatures of the photon ring can be obscured by other image structures with power on similar scales. As an example of this effect, Figure~\ref{fig::Snapshots} shows three snapshots of the GRMHD simulation from Figure~\ref{fig::GRMHD_Example} and their corresponding visibility amplitudes. Because $M \approx 9~\text{hours}$ for M87, these visibilities represent what might be obtained for a single observation of M87. While the visibilities of each snapshot show a damped radial periodicity similar to visibilities of the time-averaged image, the visibility envelope and the precise radial periodicities are affected by the transient image structures. Thus, coherent temporal averaging can potentially be used to isolate contributions of the photon ring.

\section{Additional Observational Considerations}
\label{sec::Appendix_Observational}

This section provides details on the interferometer sensitivities quoted in \S\ref{sec::Observational_Prospects} and discusses additional considerations for detecting the photon ring.

The sensitivity of an interferometer is a baseline-dependent quantity, naturally expressed through the standard deviation $\sigma$ of its complex thermal noise. This noise is independent of the source details and depends only on the geometric mean of the station sensitivities quantified via their system equivalent flux densities (SEFDs), the  integration time $\tau$, and the averaged bandwidth $\Delta\nu$:
\begin{align}
	\label{eq::sigma}
	\sigma=\frac{1}{\eta_{\rm Q}}\sqrt{\frac{{\rm SEFD}_1{\rm SEFD}_2}{2\Delta\nu\tau}}.
\end{align}
Here, $\eta_{\rm Q}$ is a correction factor of order unity caused by the quantization of received emission (for two-bit quantization, $\eta_{\rm Q}\approx0.88$). For details, see \citet{TMS}.

The SEFDs depend only on the effective area $A_{\rm eff}$ and the effective system temperature $T_{\rm sys}$ of each telescope, ${\rm SEFD}=2k_{\rm B}T_{\rm sys}/A_{\rm eff}$. The system temperature depends not only on the noise temperature of the receiver, but is also heavily affected by the sky noise and opacity. For instance, ALMA uses superconducting tunnel junction mixers, achieving a receiver noise temperature of only $230\,{\rm K}$ at 950\,GHz and 110\,K at 690\,GHz, giving SEFDs of $200\,{\rm Jy}$ and $100\,{\rm Jy}$, respectively, when the ALMA array is coherently phased. However, the effective SEFDs in good weather are approximately ${\sim}3000\,{\rm Jy}$ at 950\,GHz and ${\sim}1000\,{\rm Jy}$ at 690\,GHz. Orbiters would be unaffected by these atmospheric effects; a 4-meter orbiter with receiver noise temperature matching ALMA and an aperture efficiency of 0.5 would have a SEFD of approximately $10^5\,{\rm Jy}$ at 950\,GHz and $5\times10^4\,{\rm Jy}$ at 690\,GHz.

The limited sensitivity of a small telescope can be offset by averaging wide bandwidths, as quantified by \eqref{eq::sigma}. Wide recorded bandwidths were used by the EHT to obtain sufficient sensitivity to image M87, with data rates of 32\,gigabit\,$\text{s}^{-1}$ per site in 2017 \citepalias{PaperII}. This data rate is significantly higher than those of previous space VLBI experiments, which have utilized radio-frequency downlinks. However, optical communication can now achieve space-to-Earth data rates of ${>}100$\,gigabit\,$\text{s}^{-1}$, opening the possibility of space VLBI with extremely wide recorded bandwidths \citep[e.g.,][]{Robinson_2018}.

The maximum amount of frequency averaging is limited because different observing frequencies sample different source structures on the same physical baseline. This property also allows a single physical baseline to sample the instantaneous damped radial periodicities through observations with multiple frequencies or sufficiently wide bandwidth. A frequency displacement of $\Delta\nu$ has a corresponding radial baseline displacement of $b\Delta\nu/c$, where $b$ is the physical baseline length. Thus, the frequency separation $\Delta\nu_{d}$ corresponding to a single oscillation period in visibility amplitudes for a ring with diameter $d$ is  
\begin{align}
	\Delta\nu_d\approx155\,{\rm GHz}\pa{\frac{b}{10^4\,{\rm km}}}^{-1}\pa{\frac{d}{40\,\mu{\rm as}}}^{-1}.
\end{align}
Nevertheless, while the maximum integrated bandwidth must be significantly smaller than $\Delta\nu_{d}$ to avoid smearing the periodic signature, wider recorded bandwidths could still be coherently combined to improve sensitivity by fitting parametric models for the damped radial periodicity ($e.g.$, as in Figure~\ref{fig::Summary_Cartoon}). 

The maximum amount of temporal averaging $\tau$ is likewise geometrically limited by the rate $\ab{\dot{\mathbf{u}}}$ at which a baseline tracks through the $(u,v)$ plane \citep{TMS}. However, integration times are likely to be more severely limited, to only a few seconds or minutes, by the quality of the reference oscillators and the atmospheric phase stability. Significantly longer integrations could be possible with improved VLBI time standards, which have been explored and will also be necessary for operation at high frequencies \citep{Doeleman_2011}, or through phase-referenced observations via co-located arrays \citep{Broderick_2011} or multifrequency observations \citep{Middelberg_2005,Rioja_Dodson_2011}. For instance, multifrequency observations have been used to obtain coherence times of ${\sim}20$ minutes at 130\,GHz with the Korean VLBI Network and up to ${\sim}$hours when interleaving observations of a calibrator \citep{Rioja_2015}. 

For \sgra, blurring from scattering in the ionized interstellar medium presents a separate limitation for high resolution studies of the photon ring \citep{Zhu_2019}.\footnote{The blurring from scattering for M87 is smaller than that for \sgra\ by a factor of ${\sim}10^3$, rendering the effects negligible.} Because the blurring from interstellar scattering declines approximately with the squared observing frequency while angular resolution is proportional to frequency, every physical baseline has a minimum observing frequency imposed by the scattering,
\begin{align}
	\nu\gsim140\,{\rm GHz}\pa{\frac{b}{10^4\,{\rm km}}}.
\end{align}
Alternatively, in terms of the dimensionless baseline length, $\nu\gsim 200\,{\rm GHz}\sqrt{\frac{u}{10\,{\rm G}\lambda}}$. Thus, observations of \sgra\ with sufficient angular resolution to detect the sharp component of the photon ring ($n\geq2$) would likely require $\nu\gsim1\,{\rm THz}$.

\end{appendix}

\bibliography{bib.bib}
\bibliographystyle{apj}

\end{document}